\documentclass[times]{satauth}

\usepackage{graphicx}
\usepackage[utf8x]{inputenc}
\usepackage[english]{babel}
\usepackage{amsmath}
\usepackage{cite}
\usepackage{multirow}

\usepackage[colorlinks,bookmarksopen,bookmarksnumbered,citecolor=red,urlcolor=red]{hyperref}

\def\volumeyear{2017}

\begin{document}

\runningheads{Guidotti et al.}{LTE-based Satellite Communications in LEO Mega-Constellations}

\title{LTE-based Satellite Communications in LEO Mega-Constellations\footnote{{This work is an extension of the study presented by the Authors in ``Satellite-enabled LTE systems in LEO Constellations,'' ICC 2017 \cite{ICC_paper}.}}}

\author{Alessandro~Guidotti\affil{1}\corrauth, Alessandro~Vanelli-Coralli\affil{1}, Tommaso~Foggi\affil{2}, Giulio~Colavolpe\affil{2}, Marius~Caus\affil{3}, Joan~Bas\affil{3}, Stefano~Cioni\affil{4}, and Andrea~Modenini\affil{4}}

\address{\affilnum{1}Deptartment of Electrical, Electronic, and Information Engineering (DEI), Univ. of Bologna, 40136 Bologna, Italy \break \affilnum{2}Dipartimento di Ingegneria e Architettura, Univ. of Parma, 43124 Parma, Italy\break \affilnum{3}Centre Tecnol\'{o}gic de Telecomunicacions de Catalunya, Castelldefels, Barcelona \break \affilnum{4} European Space Agency - esa.int, ESTEC/TEC-EST, Noordwijk, The Netherlands}

\corraddr{Alessandro Guidotti, Deptartment of Electrical, Electronic, and Information Engineering (DEI), Viale del Risorgimento 2, 40136 Bologna, Italy. E-mail: a.guidotti@unibo.it}

\begin{abstract}
The integration of satellite and terrestrial networks is a promising solution for extending broadband coverage to areas not connected to a terrestrial infrastructure, as also demonstrated by recent commercial and standardisation endeavours. However, the large delays and Doppler shifts over the satellite channel pose severe technical challenges to traditional terrestrial systems, as LTE or 5G. In this paper, two architectures are proposed for a LEO mega-constellation realising a satellite-enabled LTE system, in which the on-ground LTE entity is either an eNB (Sat-eNB) or a Relay Node (Sat-RN). The impact of satellite channel impairments as large delays and Doppler shifts on LTE PHY/MAC procedures is discussed and assessed. The proposed analysis shows that, while carrier spacings, Random Access, and RN attach procedures do not pose specific issues, HARQ requires substantial modifications. Moreover, advanced handover procedures will be also required due to the satellites' movement.
\end{abstract}

\maketitle

\section{Introduction}
\label{sec:Introduction}

Nowadays, Release 14 of 3GPP Long Term Evolution (LTE), together with its evolutions LTE-Advanced (LTE-A) and LTE-A Pro, constitutes the latest standard designed for cellular communications in terrestrial systems, providing broadband connectivity up to $100$ Mbps in downlink \cite{3GPP36_201, 3GPP36_300}. This is made possible by both advanced physical layer (PHY) techniques, \emph{e.g.}, Carrier Aggregation (CA) or more aggressive Multiple Input Multiple Output (MIMO) antenna systems, and more performing architecture solutions, \emph{e.g.}, high-speed backhaul infrastructures through fiber-optic communications or Relay Nodes (RNs). However, due to the large deployment costs, in remote and rural areas, traditional backhauling solutions are not profitable for network operators due to the limited number of users that might be served, leading to an unacceptable delay in the return of their investment. Moreover, in emergency situations, terrestrial infrastructures are often destroyed or temporarily disabled, thus totally isolating the area hit by a natural disaster or a terroristic attack when connectivity is most needed, \emph{e.g.}, in order to coordinate first rescue teams.

In this context, Satellite Communication (SatCom) systems, thanks to their inherently large footprint, provide a valuable and cost-effective solution to complement and extend terrestrial networks, not only in rural areas and mission critical situations, which are critical environments as outlined above, but also for traffic off-loading in densely populated areas. Today, High Throughput Satellites (HTS) provide large capacity connectivity at reduced costs through aggressive frequency reuse schemes and multi-spot beam technology \cite{HTS_1,HTS_2}. This significant evolution is also based on valuable innovations in terms of on-board and on-ground signal processing and interference management techniques that provide means to obtain improved connectivity and flexibility \cite{IntMgmTech1, IntMgmTech2}. In addition, both Geostationary Earth Orbit (GEO) and non-GEO systems are deployed to serve air, maritime, and remote land areas in L-, S-, and Ka-bands to mobile and fixed terminals and backhaul services are also available in Ka-/Ku-bands \cite{Evans2014}. In this context, the integration of terrestrial systems with GEO satellites would result in a harmonised system providing large throughput with global coverage. Resource allocation aspects for multicast transmissions and TCP protocol performance in a GEO LTE-based satellite system have been analysed in the literature \cite{Intro1,Intro2,Intro3}, providing valuable solutions to the identified issues. The applicability of 3GPP LTE to mobile satellite systems has also been addressed in order to counteract satellite specific propagation conditions by means of a novel inter-Transmission Time Interval (TTI) interleaving technique \cite{Intro_Unibo}, which allows to disrupt the channel correlation in slowly time-varying channels. However, an even increasing attention is being directed towards Low Earth Orbit (LEO) systems and, in particular, to mega-constellations, \emph{i.e.}, systems in which hundreds of satellites are deployed, as also demonstrated by several recent commercial endeavours and funded projects \cite{Intro4}. The trend towards the integration of terrestrial and satellite systems is further confirmed by 3GPP standardisation activities for future 5G wireless systems. In particular, during the last months, a new Study Item has been issued on Non-Terrestrial Networks \cite{3GPP_5GSAT1,3GPP_SAT_WI, 3GPP_5GSAT2}, which, for the moment being, is mainly focused on the definition of deployment scenarios, system and architecture parameters, adaptation of 3GPP channel models to non-terrestrial communications, and identification of the key impact areas for the 5G air interface. Initial analyses on the feasibility of 5G-based systems in an integrated satellite-terrestrial network, and in particular on the waveform and PHY/MAC procedures, are currently being developed in the literature as well \cite{Globecom_Unibo}.

In this paper, we focus on a mega-constellation of LEO satellites deployed in Ku-band and aimed at providing LTE broadband services to areas that are not connected to a terrestrial infrastructure, \emph{i.e.}, backhauling scenarios. In Ku-band, wide bandwidths are available with respect to L-/S- bands and, thus, broadband services can be provided, while in other more limited bands low data rates or machine-type communications might be the only option. Each satellite in the considered mega-constellation serves several satellite-enabled network entities deployed on the ground that provide the radio access link to the UEs. Depending on the type of satellite-enabled entity, two architectures will be proposed. The aim of this work is to analyse and assess the impact of typical satellite channel impairments, specifically large Doppler shifts and propagation delays, in the LTE waveform design and physical and MAC layer procedures, as well as to propose solutions to the identified technical challenges. This analysis is particularly critical since LTE systems will still be a corner-stone to future 5G systems. As a matter of facts, within 3GPP standardisation activities, several LTE-to-5G migration options are currently being discussed in terms of system architecture and in most of them the traditional LTE infrastructure and core network will be the anchor system for the first implementations of 5G systems \cite{3GPP_migration1,3GPP_migration2}.

\subsection{Paper organisation and contribution}
In this paper, we analyse the feasibility of LTE-based Satellite Communication systems by means of LEO mega-constellations. In particular, in Section~\ref{sec:SystemModel}, we describe the main assumptions in terms of the typical satellite channel impairments and propose two different system architectures, based on the network entities foreseen by LTE standards. In Section~\ref{sec:LTE}, we overview the main characteristics of the LTE waveform design and physical and MAC layer procedures, \emph{i.e.}, Random Access, Hybrid Automatic Repeat reQuest (HARQ), attach procedure. These aspects are then carefully discussed in Section~\ref{sec:Challenge_Solutions} with respect to the channel impairments previously identified, and specifically large Doppler shifts and propagation delays. Several solutions to the identified technical challenges are also provided. Finally, Section~\ref{sec:Conclusions} concludes this paper.

\section{System Model}
\label{sec:SystemModel}
LEO satellites are typically deployed between $300$ and $2000$ km from Earth, which makes them particularly effective and efficient to extend and complement terrestrial networks, thanks to the reduced path loss and propagation delays with respect to GEO systems. However, Doppler rates will be significantly larger due to the high angular velocity that the satellites have on their lower orbits. Furthermore, in the following, if not otherwise specified, we assume to operate in Ku-band, in which significant signal degradations are experienced due to rain and cloud attenuation. These are even more pronounced in LEO systems because of the low elevation angle at which satellites are seen for a non-negligible percentage of time. However, the large variability in the received signal strength is reduced, and thus the throughput increased, thanks to the use of Adaptive Coding and Modulation (ACM) \cite{Propagation}.

In this paper, we consider a LEO mega-constellation satellite system aimed at providing backhaul connectivity to several on-ground network entities, which serve as interfaces between the terrestrial and satellite networks. In particular, these entities will have a two-fold purpose: on the one hand, they will serve as Base Stations, providing the radio access link towards the User Equipments (UEs) within their coverage, and, on the other hand, they will also aggregate (split) and send (receive) the information of the UEs towards (from) the LEO satellites providing the backhaul connection towards the Core Network. As we will discuss in the next paragraph, depending on the type of network functionalities of these elements, referred to as \emph{satellite-enabled network entities} in the following, different system architecture can be envisaged.

\subsection{Architecture}
\begin{figure}[t!]
\centering
\includegraphics[width=0.5\columnwidth]{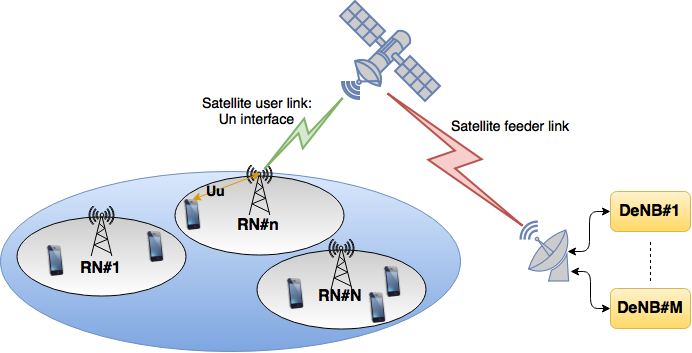}
\caption{Option A: Sat-RN architecture.}
\label{fig:Sat-RN_Arch}
\end{figure}

\begin{figure}[t!]
\centering
\includegraphics[width=0.5\columnwidth]{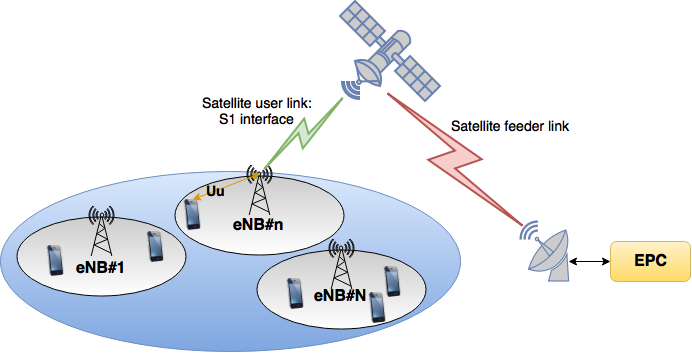}
\caption{Option B: Sat-eNB architecture.}
\label{fig:Sat-eNB_Arch}
\end{figure}

In the following, two different options for the integration of a mega-constellation of LEO satellites with LTE terrestrial networks are discussed. Both the proposed architectures are based on the following network elements and assumptions: i) on the ground segment, $N$ satellite-enabled network entities providing a traditional LTE access link, \emph{i.e.}, by means of the LTE Uu Air Interface \cite{3GPP36_300,3GPP36_401}, to the served User Equipments (UEs) are deployed in the coverage area; ii) the satellite-enabled network entities are connected to the LEO mega-constellations satellites by means of an Air Interface that depends on the specific architecture option; iii) the satellites are assumed to be transparent and to provide a backhaul connection to the satellite-enabled entities in their coverage area; iv) the system Gateway (GW) is connected to the LEO mega-constellation satellites through ideal feeder links, providing access to the LTE core network, \emph{i.e.}, Evolved Packet Core (EPC); and v) we assume a Frequency Division Duplexing (FDD) framing structure for all signals, since Time Division Duplexing (TDD) does not provide an efficient solution due to the large delays involved in satellite communications. Depending on the type of on-ground satellite-enabled entity, two different architectures, shown in Figures \ref{fig:Sat-RN_Arch} and \ref{fig:Sat-eNB_Arch}, are proposed in this paper and detailed in the following sections. It is worthwhile highlighting that, by assuming that the radio access link is always provided by means of a LTE Uu Air Interface, no modification will be required to the UEs, to which the satellite backhauling operations will be transparent.

\subsection{Option A: Sat-RN}
In the first option, we exploit the concept of Relay Nodes (RNs), that have been introduced in the LTE architecture in order to achieve a higher efficiency in network planning by means of smaller cells and heterogeneous network deployments \cite{3GPP36_300, 3GPP36_401,3GPP36_806}. A RN is a low-power Base Station wirelessly connected to a Donor eNode-B (DeNB), which provides then the connection to the core network, and it is significantly different from a typical repeater, which only amplifies and re-retransmits the received signal. As a matter of facts, RNs receive, demodulate, and decode the received data, also applying error correction algorithms, and then re-construct the signal to be transmitted to the users in their coverage area. In terms of Air Interfaces and operation mode, RNs are connected to the UEs in their coverage area by the traditional Uu Air Interface, while the Un Air Interface is exploited on the backhaul link towards the DeNB. From the network functionality point of view, RNs terminate the Uu, S1 (Air Interface between eNBs and Evolved Packet Core network), and X2 (Air Interface between eNBs) protocols up to Layer 3 \cite{3GPP36_401,3GPP36_806}. Thus, as a matter of facts, the RN acts as an eNB from the perspective of the UEs, while it acts as a simple UE from the DeNB point of view. Furthermore, LTE specifications clearly state that the Un Air Interface uses the same radio protocols and procedures as the Uu, with the only difference being in the Radio Frequency characteristics and minimum performance requirements \cite{3GPP36_300,3GPP36_116}. 

Based on the above, in option A, it is assumed that the satellite-enabled network entity in the ground segment is a LTE Relay Node and the resulting system architecture is shown in Figure~\ref{fig:Sat-RN_Arch} and will be referred to as \emph{Sat-RN} architecture. Taking into account the above discussion on how RNs operate in the LTE system, two critical aspects are worth being highlighted in this architecture: i) on the backhaul link (RN-to-DeNB), the RN is connected to a DeNB by means of the Un Air Interface, which is a modified version of the traditional Uu interface, while on the access link (UE-to-RN) the UEs communicate to the RN through the traditional Uu interface; and ii) the RN supports eNB functionalities, \emph{i.e.}, it terminates the Uu, S1 (Air Interface between eNBs and the EPC), and X2 (Air Interface between eNBs) protocols up to Layer 3. Since for the Un Air Interface the only difference with respect to the Uu interface is in the Radio Frequency characteristics and minimum performance requirements, for both the backhaul and access links, the system adopts Orthogonal Frequency Division Multiplexing (OFDM) and Single Carrier-Frequency Division Multiple Access (SC-FDMA) in the downlink and uplink, respectively. Finally, as shown in Figure~\ref{fig:Sat-RN_Arch} and described in the LTE standard, a single Donor eNB can manage more than one RN, \emph{i.e.}, $M\leq N$, where $M$ and $N$ denote the number of DeNBs and RNs, respectively.

\subsection{Option B: Sat-eNB}
The second option that can be envisaged for the considered system is based on the traditional LTE macro-cell architecture, \emph{i.e.}, on the deployment on eNBs. In LTE, eNBs have the capability to operate with several air interfaces depending on the type of communication. In particular, the X2 air interface allows different eNBs to interconnect and exchange information in order to improve the overall network management efficiency, while the S1 air interface is defined for the connection between eNBs and the EPC. Since in this work we aim at discussing the backhaul link, we will assume that no interconnection among on-ground satellite-enabled entities is present and, thus, focus on the S1 air interface. As usual in LTE, this air interface is split in two parts, one for the Control Plane (CP) and one for the User Plane (UP), and it is based on a full IP protocol stack. The critical aspect to be mentioned is that in LTE specifications it is clearly stated that the S1 air interface is open \cite{3GPP36_116,3GPP36_410}. In particular, the S1 interface can be of any type as long as it supports the exchange of signalling between the eNB and the EPC. This choice in the LTE standards was driven by the aim at facilitating: i) the interconnection of eNBs with Mobility Management Entities (MMEs) supplied by different manufacturers; and ii) the separation of the of S1 interface Radio Network functionality and Transport Network functionality to facilitate introduction of future technology.

Based on the above, in the second option the satellite-enabled network entity is a traditional LTE eNB, as depicted in Figure~\ref{fig:Sat-eNB_Arch}, and we will refer to this deployment as \emph{Sat-eNB} architecture. Since, differently from the Sat-RN case, we have a traditional eNB providing on-ground connectivity to the UEs, the satellite user link adopts the S1 interface. The S1 interface is an open interface, \emph{i.e.}, it can be any radio interface as long as a few (signalling) mandatory requirements are met \cite{3GPP36_116,3GPP36_410}. Thus, for the Sat-eNB architecture, the interface can be implemented as a Uu, Un, or a SatCom specific-designed (\emph{e.g.}, DVB-S2X \cite{DVBS2X}) air interface as long as the mandatory requirements for S1 interfaces are met.

\subsection{Satellite Channel}
In the following, we focus on the Sat-RN architecture option and assess the feasibility of LTE in the considered LEO mega-constellation system, since the Sat-eNB option has an intrinsic standard-compliant solution for optimised transmissions on the satellite backhaul link. In particular, the main technical challenges in implementing the Un air interface over the satellite link will be discussed by focusing on the waveform design and PHY/MAC layer procedures. As a matter of facts, these aspects might be strongly impacted by typical satellite channel impairments as large Doppler shifts and propagation delays, and, thus, a proper analysis is critical to the realisation of LTE-based SatCom. With respect to the propagation delay, the Round Trip Time (RTT) is the considered metric, which is defined as twice the propagation delay between the transmitter and the receiver, under the assumption that the propagation delay is the same for both uplink and downlink, plus the signal processing time. Since we are in a Satellite Communication context, we can assume that the signal processing time is negligible with respect to the propagation delay and, thus, approximate the RTT with twice the propagation delay. As for the Doppler shift, it consists in the change in the carrier frequency due to the relative motion between the satellite and the user terminal. The maximum Doppler shift can be computed for a given satellite altitude and carrier frequency and minimum elevation angle at which the satellite is seen from the terminal \cite{Doppler_LEO}.

In the following, the Doppler shift tolerated by the receiver and the impact of the large RTT on LTE protocols and waveforms will be discussed. It shall be noticed that, in the considered scenario, we have both the access link (UE-to-RN) and the backhaul link (RN-to-DeNB). However, only the backhaul link will be considered for such analysis, as the access link is involved in a traditional LTE cell for which no modification to procedures or requirements shall be introduced, as will also be highlighted in the next sections. Besides general considerations, we assume a LEO mega-constellation operating in Ku-band at an altitude of $h = 1200$ km, a beam size of approximately $320$ km, and a minimum elevation angle of $45^{\circ}$.
\begin{figure}[t!]
\centering
\includegraphics[width=0.4\columnwidth]{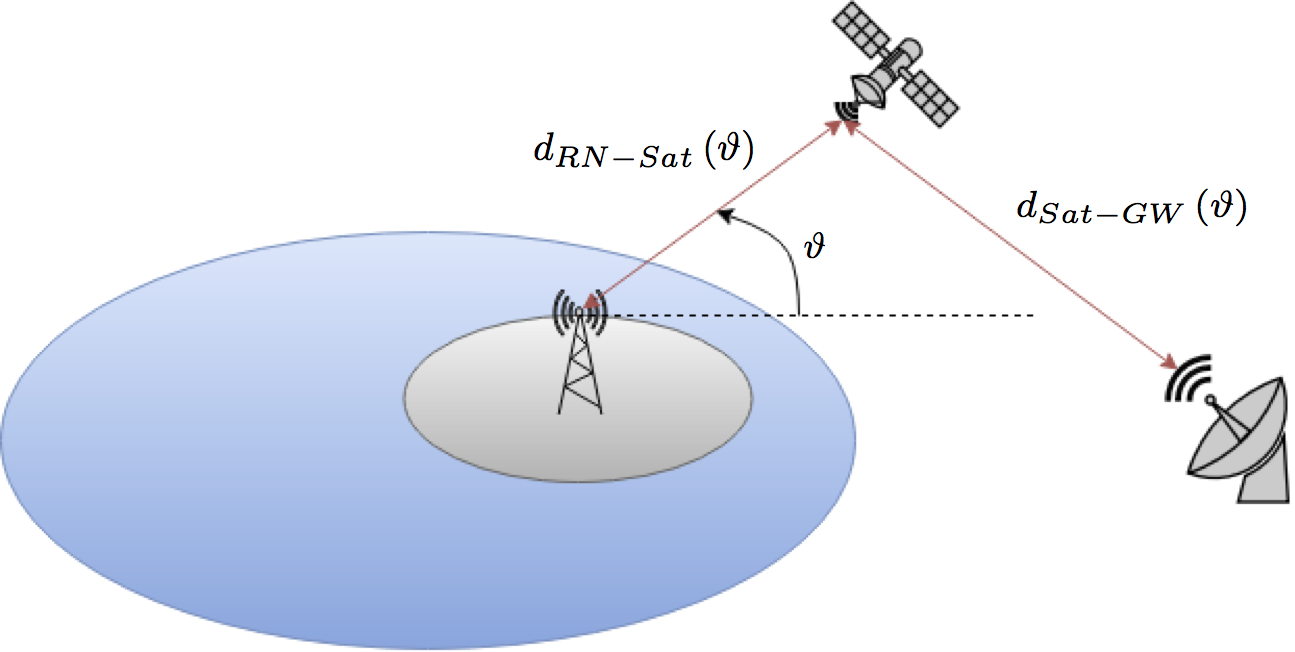}
\caption{Reference geometry for the Round Trip Time computation in the Sat-RN architecture.}
\label{fig:Delay_figure}
\end{figure}
\subsubsection{Delay}
\label{sec:Delay}
The propagation delay is a critical issue, since it yields a misalignment between uplink and downlink frames and it also affects MAC layer protocols. This is especially relevant in SatCom, where distances are very large when compared to typical distances in terrestrial networks, even for Low Earth Orbit systems as the one at hand. As previously highlighted, we can assume that the RTT delay is given by twice the one way propagation delay, since the propagation delay is significantly larger than the signal processing time in SatCom \cite{Maral}. In the considered scenario, the one-way propagation delay is given by the combination of the delay on the RN-to-Satellite link and the delay on the Satellite-to-GW path, which are a function of the elevation angle as follows, see Figure~\ref{fig:Delay_figure}:
\begin{equation}
\label{eq:RTT_OneWay}
\begin{split}
    T_{1way} &= T_{RN-Sat}\left(\vartheta_1\right) + T_{Sat-GW}\left(\vartheta_2\right) = \frac{d_{RN-Sat}\left(\vartheta_1\right) + d_{Sat-GW}\left(\vartheta_2\right)}{c} \overset{(a)}{=} \\
    & \overset{(a)}{=} \frac{R_E\left[ \sqrt{\frac{{\left(h+R_E\right)}^2}{R_E^2}-\cos^2\vartheta_1} - \sin\vartheta_1\right]}{c} + \frac{R_E\left[ \sqrt{\frac{{\left(h+R_E\right)}^2}{R_E^2}-\cos^2\vartheta_2} - \sin\vartheta_2\right]}{c}
\end{split}
\end{equation}
where: i) $R_E$ is the Earth radius; ii) $c$ is the speed of light; iii) in $(a)$ we exploited the slant range formula; and iv) $\vartheta_1$ and $\vartheta_2$ are the elevation angles for the RN and the GW, respectively. From the propagation delay point of view, the worst-case scenario is obtained by computing (\ref{eq:RTT_OneWay}) with $\vartheta_1=\vartheta_2=\vartheta=45^{\circ}$, which is the minimum elevation angle assumed in this study. In this case, we obtain $d_{RN-Sat}\left(45^{\circ}\right)=d_{Sat-GW}\left(45^{\circ}\right)\approx 1580$ km and:
\begin{equation}
\label{eq:RTT_TwoWay}
    RTT \approx T_{2way} = 2T_{1way}\approx 21.06 \ \mathrm{ms} 
\end{equation}
In the LTE standard, the propagation delay between the UEs and the serving eNB has an impact on the so called Timing Advance (TA). In a traditional LTE cell, a UE far from the eNB experiences a larger propagation delay and, thus, its uplink transmission is received later by the eNB, with respect to a UE closer to the eNB. In order to cope with such time misalignment, LTE introduces a Timing Advance, which allows to adjust the transmission time in order to synchronise uplink and downlink frames at the eNB. Depending on the specific configuration, the maximum delays in an LTE cell are included in the range $[133.33, 688.021]$ $\mu$s and the maximum timing advance allowed by the protocol is $T_{AD} = 0.6667$ ms \cite{3GPP36_211}. This is defined as $T_{AD} = 16\cdot T_{A} \cdot T_{S}$, where $T_{S} = 1/(2048\cdot 15000)$ s is the sampling time and TA is a parameter included in the range $[0,1282]$ \cite{3GPP36_213}. As it can be noticed from (\ref{eq:RTT_TwoWay}), the RTT delay in the considered LEO mega-constellation is much larger than the maximum $T_{AD}$ foreseen by the LTE standard. However, no modification to the TA is required in the considered Sat-RN system since: i) in the radio access link, we have a traditional LTE cell for which no modifications are needed, because we assumed a beam coverage with a size of approximately $320$ km, in which several RNs are deployed to provide the access link to UEs and each RN has, thus, a coverage area that is within the maximum coverage foreseen by LTE (approximately $100$ km); and ii) on the backhaul link, the RN gathers information from all of the UEs in its cell, aggregates the data, and transmits it to the satellite, which implies that all uplink transmissions are coming from the same entity. Although no modifications are needed to the TA, the large RTT in a satellite system will have a strong impact on the PHY and MAC layers, \emph{e.g.}, the Random Access and the Hybrid Automatic Repeat reQuest (HARQ) procedures, as will be discussed in the following sections.

\begin{figure}[t!]
\centering
\includegraphics[width=0.4\columnwidth]{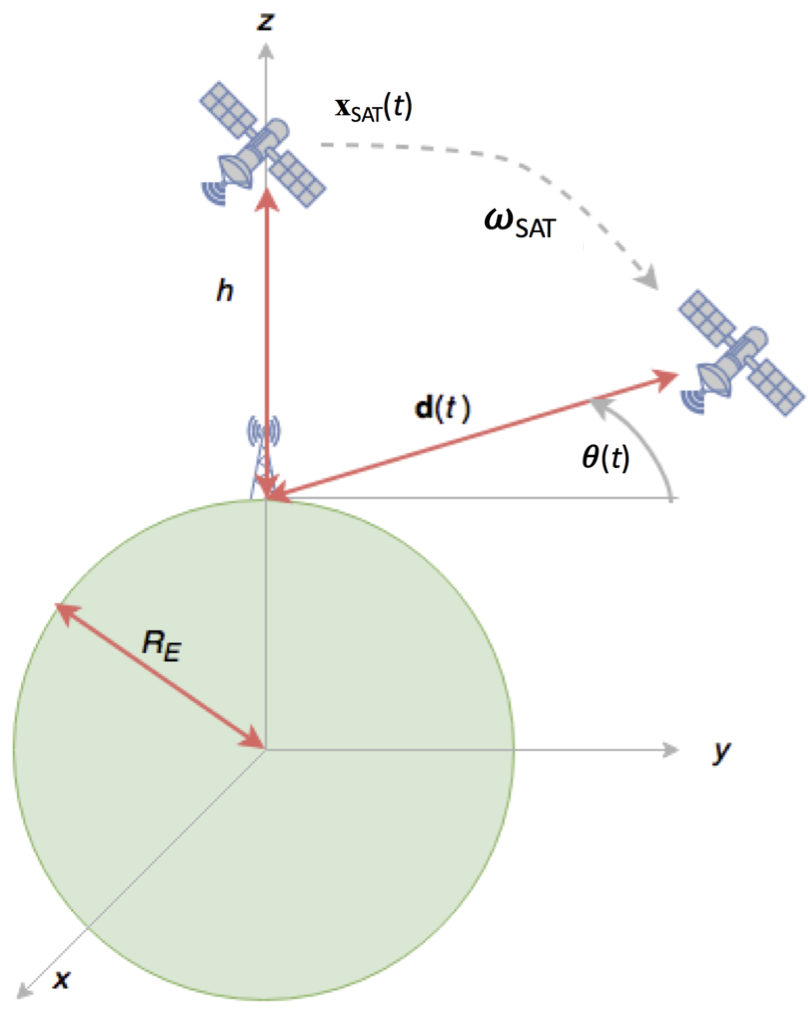}
\caption{Reference geometry for the computation of the Doppler shift.}
\label{fig:DopplerGeometry}
\end{figure}
\subsubsection{Doppler}
LTE was designed to ensure the communication even from high speed trains, traveling at around $v=500$ km/h. Assuming a carrier frequency $f_\text{c}=2$ GHz, and by reminding that the Doppler shift can be computed as $f_{\text{d}}=(v\cdot f_{\text{c}})/c$, where $v$ is the relative speed between the transmitter and the receiver, it is possible to obtain a maximum bearable Doppler shift $f_{\text{d}}= 950$ Hz, and, according to the Nyquist sampling theorem, the maximum sampling period to correctly estimate the channel is $0.5$ ms.

The system described in the previous section entails different considerations for the access (UE-to-RN) and backhaul (RN-to-DeNB through satellite) link. In the first case, the situation is the same as for terrestrial LTE, whereas in the second case, the maximum Doppler shift from LTE specifications is significantly exceeded, as the satellite orbital speed induces a very large Doppler shift (typically, the LEO satellites orbital speed is in the order of several km/s). In more detail, the Doppler shift is equal to zero when the satellite is at the zenith, whereas at lower elevation angles, larger Doppler shift values, which have to be coped with, arise. Figure~\ref{fig:DopplerGeometry} shows the fundamental parameters which will be used to compute these critical values. It shall be noticed that, due to Earth's rotation, the Doppler shift might deviate from the value computed by means of the great-circle arc. However, at LEO altitudes, this approximation still provides accurate results. Based on this geometry, the Doppler shift experienced by a generic RN can be expressed as a function of time as follows:

\begin{equation}
\label{eq_Doppler}
    f_{\text{d}}(t) = \frac{f_\text{0}}{c}\frac{\textbf{d}(t)}{\left|\textbf{d}(t)\right|}\frac{\partial \textbf{x}_\text{SAT}(t)}{\partial t}
\end{equation}
where $f_\text{0}$ is the carrier frequency, $\textbf{d}(t)$ is the distance vector between the satellite and the RN, and $\textbf{x}_\text{SAT}(t)$ is the vector of the satellite position:
\begin{equation}
\label{eq:Sat_position}
    \textbf{x}_{\text{SAT}}(t)=[0,(R_{\text{E}}+h)\text{sin}(\omega_{\text{SAT}}t),(R_{\text{E}}+h)\text{cos}(\omega_{\text{SAT}}t)]^T
\end{equation}
Referring to the geometry provided in Fig.~\ref{fig:DopplerGeometry}, we have:
\begin{equation}
\label{eq:RN_Sat_distance}
    \textbf{d}(t) = [0,(R_\text{E}+h) \text{sin}(\omega_{\text{SAT}}t),(R_\text{E}+h) \text{cos}(\omega_{\text{SAT}}t)-R_\text{E}]^T
\end{equation}
where $\omega_{SAT}$ is the satellite angular velocity. To express the Doppler shift as a function of the satellite elevation angle, we need to write $\omega_{SAT}t$ as a function of $\left|\textbf{d}(t)\right|$, $\theta(t)$, $h$, and $R_E$. Such relation can be easily obtained as follows:
\begin{equation}
\label{eq:Distance_Pitagora}
    \text{cos}(\omega_{\text{SAT}}t) = \frac{R_E+\left|\textbf{d}(t)\right|\text{sin}(\omega_{\text{SAT}}t)}{R_E+h}
\end{equation}
By plugging (\ref{eq:Sat_position}), (\ref{eq:RN_Sat_distance}), and (\ref{eq:Distance_Pitagora}) into (\ref{eq_Doppler}) we get
\begin{equation}
    \label{eq4}
    f_d(t) = \frac{f_0R_E\omega_{SAT}}{c}\sqrt{ \frac{h^2 + 2R_Eh - 2R_E\left|\textbf{d}(t)\right| \sin\left(\vartheta (t)\right) - {\left|\textbf{d}(t)\right|}^2 {\sin\left(\vartheta (t)\right)}^2 }{h^2 + 2R_Eh - 2R_E\left|\textbf{d}(t)\right|\sin\left(\vartheta (t)\right)} }
\end{equation}
Recalling the law of cosines, we can relate the sides of the triangle defined by the center of the Earth, $\textbf{x}_{\text{SAT}}(t)$ and the RN position as
\begin{equation}
    \left(R_E+h\right)^2 - R_E^2 - {\left|\textbf{d}(t)\right|}^2 = 2R_E\left|\textbf{d}(t)\right|\sin(\vartheta(t))
\label{eq5}\end{equation}
By substituting (\ref{eq5}) into (\ref{eq4}), we finally obtain the following closed-form expression of the Doppler shift as a function of the elevation angle:
\begin{equation}
\label{eq:Doppler}
    f_\text{d}(t)=\frac{f_\text{0} \omega_{\text{SAT}} R_\text{E} \text{cos}(\vartheta(t))}{c}
\end{equation}
where $\omega_{\text{SAT}}=\sqrt{GM_E/(R_E+h)^3}$, $G=6,6710^{-11}$ $\textrm{N}\cdot \textrm{m}^2/\textrm{kg}^2$ is the Gravitational constant, and $M_\text{E}=5,98\cdot 10^{24}$ kg is the Earth mass. Notice that the orbital speed $v_{\text{SAT}}=\omega_{\text{SAT}}(R_E+h)$ is equal to 7.255 km/s, much higher than the 500 km/h envisaged by LTE specifications. In the considered scenario, if we assume a carrier frequency in Ku-band, \emph{i.e.}, in the range $11$ GHz $\leq f_\text{0} \leq 14$ GHz, a maximum Doppler shift in the range $158$ kHz $\leq f_{\text{d}}\leq 201$ kHz is obtained. Thus, the Doppler shift in the satellite channel is increased by a factor $x$, with $166<x<211$, with respect to the maximum Doppler shift experienced in LTE.

\section{LTE waveform and PHY/MAC procedures}
\label{sec:LTE}
In this section, we provide a quick overview of the main aspects related to the LTE standard and highlight the most critical technical challenges introduces by the satellite channel impairments reviewed in the previous section. In particular, we will focus on the subcarrier spacing in the LTE waveform and on several PHY/MAC layer procedures, \emph{i.e.}, Random Access (RA), Hybrid Automatic Repeat reQuest (HARQ), and the Relay Node attach procedure.
\subsection{Waveform}
As reported in Section \ref{sec:SystemModel}, the downlink waveform is based on OFDM, whereas the uplink is based on SC-FDMA. The transmission resources are defined in three dimensions: time, frequency, and space. The latter is measured in layers and is accessed in terms of antenna ports at the eNB. Downlink and uplink transmissions are organised into frames with $10$ ms duration, consisting of $20$ time slots with duration $0.5$ ms or, equivalently, $10$ sub-frame with a duration of $1$ ms each. In each time slot there are $6$ or $7$ OFDM (or SC-FDMA) symbols, according to what type of cyclic prefix (CP) is being used. OFDM (or SC-FDMA) symbols per subcarrier per time slot are called resource elements, so that the product of resource elements and the overall subcarriers defines a resource block. When an extended CP is implemented, the subcarrier spacing $\Delta f$ can be either set to $15$ kHz or to $7.5$ kHz, whereas for normal CP it is $15$ kHz only. Details on LTE downlink resource grid parameters are reported in Table \ref{tab:DL_param}.

\begin{table}[h]
\centering
\footnotesize
\caption{Downlink physical layer parameters.}{\footnotesize \textsuperscript{(1)}: the overall number of subcarriers is $N_{RB}^{DL}N_{sc}^{RB}+1$, but the DC is not transmitted.}
\label{tab:DL_param}
\begin{tabular}{|c|c|c|c|c|c|c|c|}
\hline
\textbf{Parameter} & \textbf{Unit} & \multicolumn{6}{c|}{\textbf{Value}}\\ \hline \hline
\multicolumn{8}{|c|}{\emph{Resource grid parameters}}\\ \hline
Transmission bandwidth & [MHz] & 1.4 & 3 & 5 & 10 & 15 & 20\\ \hline
Frame duration $T_f$ & [ms] & \multicolumn{6}{c|}{10}\\ \hline
Sub-frame duration & [ms] & \multicolumn{6}{c|}{1}\\ \hline
Time slot duration $T_{slot}$ & [ms] & \multicolumn{6}{c|}{0.5}\\ \hline
RB per time slot $N_{RB}^{DL}$ & & 6 & 15 & 25 & 50 & 75 & 100 \\ \hline 
Subcarrier spacing $\Delta f$ & [kHz] & \multicolumn{6}{c|}{15} \\ \hline
OFDM symbols per time slot $N_{symb}^{DL}$, normal CP & & \multicolumn{6}{c|}{7}\\ \hline
OFDM symbols per time slot $N_{symb}^{DL}$, extended CP & & \multicolumn{6}{c|}{6}\\ \hline
Subcarriers per RB $N_{sc}^{RB}$ & & \multicolumn{6}{c|}{12}\\ \hline
Active subcarriers per time slot $N_{RB}^{DL}N_{sc}^{RB}$ \textsuperscript{(1)} & & 72 & 180 & 300 & 600 & 900 & 1200 \\ \hline
Occupied bandwidth (excl. DC) & [MHz] & 1.08 & 2.7 & 4.5 & 9.0 & 13.5 & 18.0 \\ \hline \hline
\multicolumn{8}{|c|}{\emph{OFDM parameters}}\\ \hline
\multirow{2}{*}{Normal CP} & [$\mu$s] & \multicolumn{6}{c|}{5.21 ($l=0$), 4.69 ($l=1,\ldots,6$)}\\ \cline{2-8}
    & $T_s$ & \multicolumn{6}{c|}{160 ($l=0$), 144 ($l=1,\ldots,6$)} \\ \hline
\multirow{2}{*}{Extended CP} & [$\mu$s] & \multicolumn{6}{c|}{16.67 ($l=0,\ldots,5$)}\\ \cline{2-8}
    & $T_s$ & \multicolumn{6}{c|}{512 ($l=0,\ldots,5$)} \\
    \hline
FFT size & & 128 & 256 & 512 & 1024 & 1536 & 2048\\ \hline
Sampling rate & [MHz] & 1.92 & 3.84 & 7.68 & 15.36 & 23.04 & 30.72 \\ \hline
Samples per slot & & 960 & 1920 & 3840 & 7680 & 11520 & 15360 \\ \hline
Useful symbol duration & [$\mu$s] & 66.67 & 66.67 & 66.67 & 66.67 & 66.67 & 66.67 \\\hline
\end{tabular}
\end{table}

It is worth highlighting that LTE provides robustness to a high extent against the detrimental effects induced by the user mobility, as in the case of high speed trains. The LTE protocol envisages a set of resource elements used by the physical layer that does not carry any information from the upper layers. In fact, they are used for synchronisation and estimation purposes (\emph{e.g.}, reference signals known at both the transmitter and the receiver and used for channel estimation, to allow coherent demodulation at the receiver, and to estimate the position). However, it is required that the Doppler shift is significantly lower than the subcarrier spacing and that the channel does not experience strong variations in one time slot. From the analysis conducted in Section \ref{sec:SystemModel}, it clearly emerges that the aforementioned constraints will not be satisfied in the satellite systems under study. Consequently, the properties of primary and secondary synchronisation signals transmitted in the downlink frames could not be properly exploited to estimate the frequency misalignment between the transmitter and the receiver. In addition, reference signals embedded with the data cannot be used to estimate the channel with the desired accuracy and, thus, channel variations cannot be tracked. In the light of this discussion, potential solutions to overcome the impairments that are inherent to LEO satellite communication systems will be discussed in Section \ref{sec:WaveformSolutions}. 

\begin{figure}[t!]
\centering
\includegraphics[width=0.5\columnwidth]{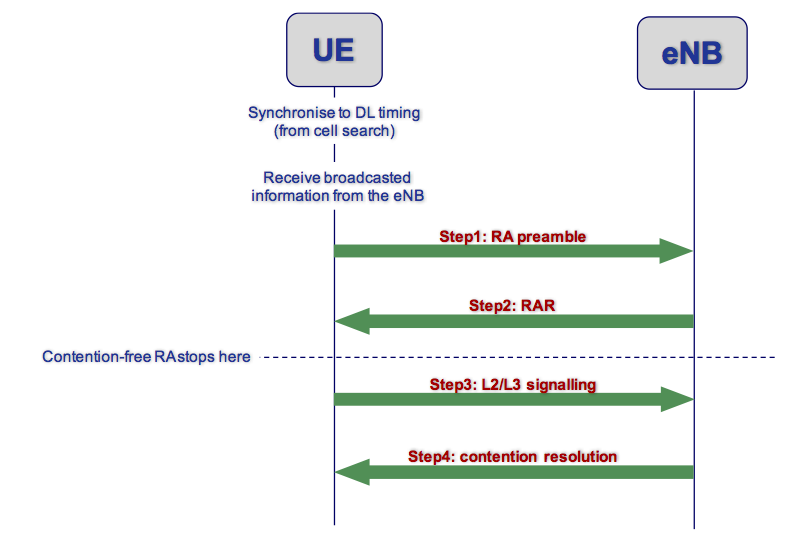}
\caption{LTE Random Access procedure: contention-based and contention-free.}
\label{fig:LTE_RA}
\end{figure}
\subsection{Random Access}
In LTE, there are two types of Random Access procedure, as shown in Figure\ref{fig:LTE_RA}: i) contention-based (non-synchronised), which is used when the UE is not synchronised or lost its synchronisation to the eNB; and ii) contention-free (synchronised), to be used when the UE is already logged to the network and needs to perform an handover procedure. Both procedures are based on the transmission of a RA preamble, to be built and transmitted according to the broadcast control information received from the eNB, which include, among the others, the maximum number of RA attempts, timer settings for the RA procedure, and the PHY resources to be used for the procedure. Since the contention-free procedure is included in the contention-based one, we focus on the latter and hereafter briefly described the four steps, in order to fully understand the impact of the satellite channel impairments in the considered system.
\paragraph{Step 1: RA preamble transmission} 
The RA preamble is randomly chosen from $64$ orthogonal sequences, composed by $0<N_{CF}<60$ sequences reserved for the contention-free approach and $N_{CB} = 64-N_{CF}$ that can be used for the contention-based procedure \cite{3GPP36_211}. The $N_{CB}$ sequences available for contention-based RA are split into two groups, so as to use one bit to inform the eNB on the expected amount of resources that will be needed. In case the UE previously performed a RA attempt, it shall use the same PHY resources and sequence group as before, with the only exception being an increased transmission power within a range specified by the eNB. It is worthwhile highlighting that the RA preamble is used at the eNB to estimate the delay between it and the UE, so as to compute the Timing Advance. Finally, together with the RA preamble, the UE sends a temporary network identifier so as to allow the eNB to recognise and refer to its RA procedure.
\paragraph{Step 2: RA response} 
Once the UE sends the RA preamble, it starts monitoring the Physical Downlink Control Channel (PDCCH) for a Random Access Response (RAR) carrying its temporary identifier. The RAR shall be received within a predefined time window starting from the sub-frame containing the end of the preamble plus $3$ sub-frames, with a duration that can be set up to $10$ ms. Since the LTE sub-frame has a fixed duration of $1$ ms, this implies that the RAR shall be received by the UE within $15$ ms after its transmission. In this phase, one of the two following outcomes is possible: i) the RAR is correctly received within the RAR time window, and thus the UE implements the TA, if any, and proceeds with Step 3; or ii) the RAR is not correct or the RAR time window expired and, thus, the UE can start a new RA procedure from Step 1, as long as it did not reach the maximum number of attempts, which can be set up to $200$ \cite{3GPP36_331}. If the maximum number of RA attempts has been reached, the RA is reported as unsuccessful to the upper layers. It shall also be noticed that this is the last step for contention-free procedures.
\paragraph{Step 3/4: Msg3 and contention-resolution}
Assuming that the UE received a correct RAR within the RAR time window, it sends a new message to the eNB, denoted as Msg3, containing the terminal identity to be used in Step 4, \emph{i.e.}, the contention-resolution phase. In this phase, HARQ is implemented and a contention timer is started at the UE as soon as Msg3 is transmitted. This timer can be set up to $64$ sub-frames, \emph{i.e.}, $64$ ms, and it is started again at each attempt to transmit the Msg3, up to $8$ tentatives. In Step 4, the contention resolution phase, there are two possibilities: i) if the UE is addressed with the correct identifier and within the contention timer, then the RA is successful and it can start transmitting and receiving information; or ii) if the identifier in the message from the eNB is not correct or the contention timer already expired, the UE can try a new Msg3 transmission or, if the maximum number of attempts has been reached, start a new RA procedure (in case new attempts are available, as discussed above).

\subsection{RN attach procedure}
In the considered system, it shall be noted that the RA procedure is performed in two separate situations. First of all, within each on-ground LTE cell, the UEs try to connect to the RN by means of a traditional contention-based or contention-free procedure, since the RNs are seen as normal eNBs, as discussed above. Apart from this traditional RA, for which the technical challenges are discussed in the next section, the RA procedure shall also be performed by the RN when it first connects to the network, which is the RN attach procedure. This procedure is performed in two phases, as described below \cite{3GPP36_300}.
\paragraph{Phase I}
Before it is allowed to operate as a normal relay, the RN, at start-up, attaches to the network as a normal UE to any eNB in the network (\emph{i.e.}, it is not required in this step that it contacts a Donor eNB), so as to receive the required information and configurations to operate as a relay. Thus, there are three relevant aspects to be noted: i) the RN sends no information on being a relay in the Radio Resource Control (RRC) layer signalling; ii) the eNB treats the RN as an UE; and iii) the Mobility Management Entity (MME) does not perform any RN-related operation. The Home Subscriber Station (HSS) informs the MME that the node is a RN acting as a UE and the RN receives the configuration parameters, as well as a list of available DeNBs, as User Plane traffic. Finally, the RN detaches from the network as an UE and moves to phase II.
\paragraph{Phase II}
The RN reconfigures its operating parameters and attaches to one of the DeNBs belonging to the list received in phase I. It shall be noted that the DeNB must be aware of the MMEs support for RNs and, differently from phase I, the RN clearly specifies that it is operating as a relay. The RN can then finalise its re-configuration and start accepting UEs within its coverage.

\subsection{HARQ}
The main purpose of implementing an HARQ protocol consists in improving the link reliability. If the packet is correctly decoded by the receiver, then a positive acknowledgment (ACK) is sent to the transmitter. Otherwise, when the cyclic redundancy check (CRC) fails, a negative confirmation answer is sent, \emph{i.e.}, NACK. In the LTE MAC, up to $8$ HARQ parallel processes are present in the HARQ entity at MAC layer so as to better exploit the available resources and they are based on a Stop-And-Wait (SAW) protocol. Since the transmission is performed in $1$ ms, there is a $8$ ms periodicity in the operations performed by each HARQ process. Similarly to the RA, two different HARQ processes are implemented in LTE: i) in the downlink, an adaptive (the transmission attributes are adaptively reconfigured to the channel conditions) asynchronous (retransmissions can happen at any time, thus requiring an HARQ identifier to distinguish among the different HARQ parallel processes) scheme is implemented; and ii) in the uplink, the process is synchronous, thus not requiring the transmission of a HARQ identifier, and can be either adaptive or non-adaptive (in this case, the transmission parameters are modified on a predetermined basis). 

\paragraph{Downlink HARQ}
With respect to the downlink procedure, initially the UE does not have any knowledge on HARQ processes and, thus, it shall receive information from the eNB or RN on the broadcast control channels. This provides the UE the HARQ identifier (required in asynchronous schemes), Modulation and Coding Scheme (MCS), and Redundancy Version (RV), \emph{i.e.}, the redundancy used in the coding scheme. Since this approach is asynchronous, the HARQ ID must be sent at each (re-)transmission. Moreover, being the retransmission non-adaptive, the RV and MCS follows a predefined incremental redundancy sequence. Due to the asynchronous scheme, the ACK/NACK for the data sent by the eNB or RN in the generic $n$-th sub-frame must be triggered by an uplink grant for the UE for ACK/NACK reporting in subframe $n - 4$.
\paragraph{Uplink HARQ}
In the uplink, an adaptive/non-adaptive synchronous scheme is implemented. Since the HARQ is synchronous, no HARQ ID is required: the round-robin operation of the $8$ HARQ processes allows to implicitly identify the addressed process. In the adaptive case, the PHY resources and MCS to be used are modified at each re-transmission so as to adapt to the current channel conditions, while, in the non-adaptive case, the same set of PHY resources and MCS is used for retransmissions and the only required parameter is a flag denoting whether a new data block is being transmitted or it is a re-transmission. The RV, as for the downlink case, follows a pre-defined incremental redundancy. The maximum number of re-transmissions is set to $1, \ldots, 8, 10, 12, 16, 20, 24, 28$ (with $5$ being the default parameter) and the ACK/NACK shall be transmitted (eNB-to-UE) $4$ ms after the data transmission (UE-to-eNB). When taking into account RNs, due to the presence of broadcast signalling and synchronisation signals in the PHY resources, a $4$ ms periodicity is not feasible in the uplink. In particular, the permission to retransmit the message, in some configurations, cannot be received $4$ ms in advance. Thus, only configurations with a basic periodicity equal to $8$ ms are implemented and up to $6$ HARQ parallel processes are present, so as to minimise the delays.

\begin{figure}[h!]
\centering
\includegraphics[width=0.4\columnwidth]{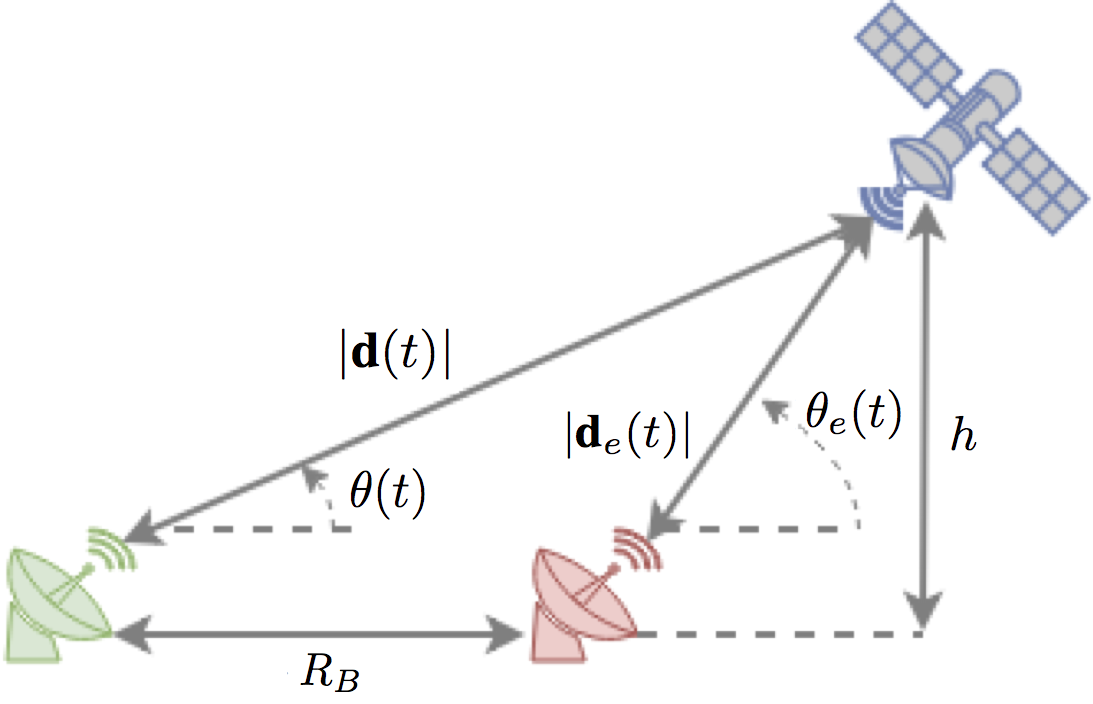}
\caption{RN position displacement for the computation of the residual Doppler shift.}
\label{fig:true_dopp}
\end{figure}
\section{Technical challenges and proposed solutions}
\label{sec:Challenge_Solutions}
\subsection{Waveform}
\label{sec:WaveformSolutions}
\begin{figure}[h!]
\centering
\includegraphics[width=0.5\columnwidth]{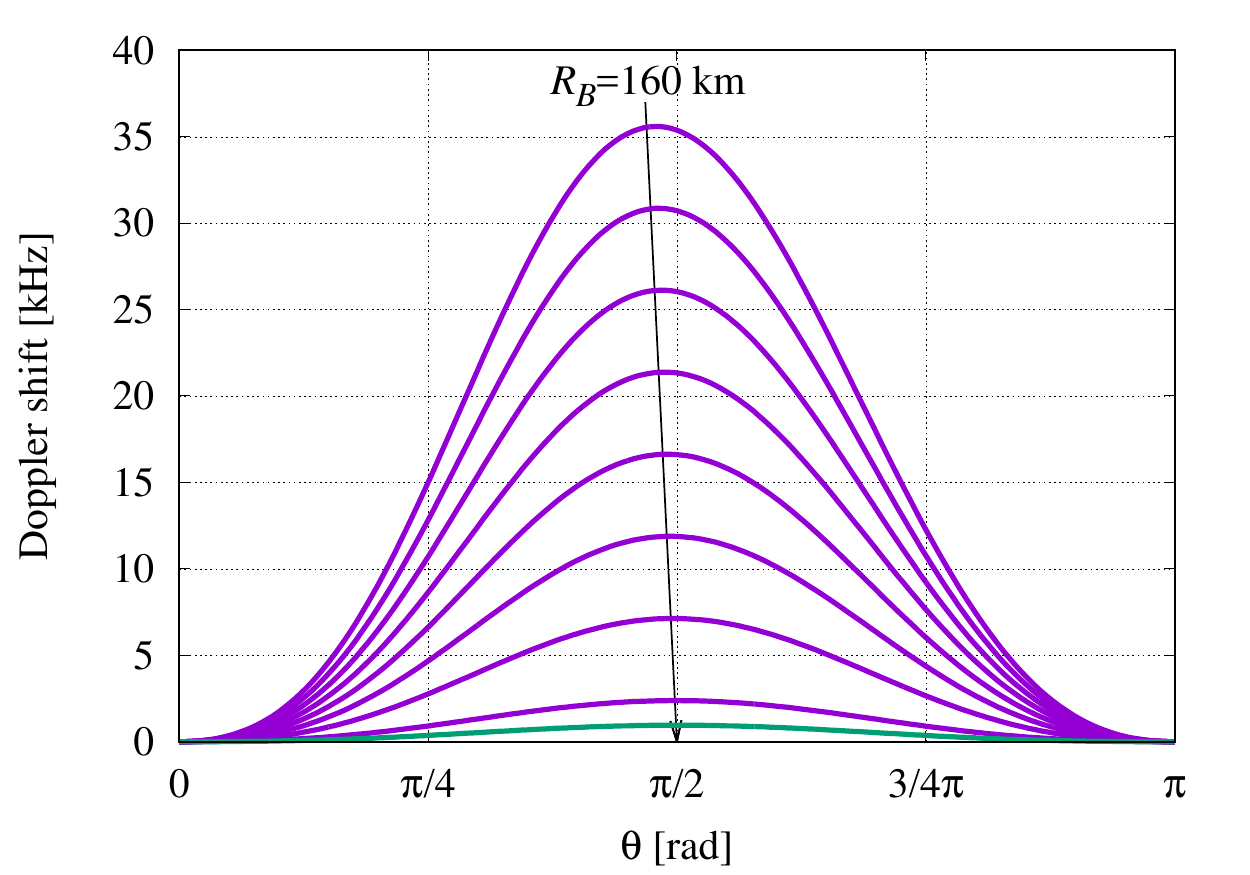}
\caption{Residual Doppler shift as a function of $\theta$, with $R_B$ decrease step equal to 20 km.}
\label{fig:res_dopp}
\end{figure}

As outlined in Section~\ref{sec:SystemModel}, the Doppler shift on the backhaul link is increased by a factor $166 < x < 211$ with respect to the maximum Doppler envisaged by LTE specifications and, thus, some modifications are required to implement it in the Sat-RN architecture. As a matter of fact, the Doppler shift could be compensated with acceptable accuracy by equipping the RN with a Global Navigation Satellite System (GNSS) receiver and providing also the trajectory of the satellite. Clearly, an estimation error in the computation of the relative position between the RN and the satellite would entail a residual Doppler shift on the signal. From geometrical considerations, the difference between the compensated and the actual Doppler shift can be computed, as explained in the following. 
Let us assume that the position $\textbf{x}_{\text{SAT}}(t)$ of the satellite is known. Then, at a given time instant, the exact distance $\left|\textbf{d}(t)\right|$ and elevation angle $\vartheta(t)$ can be computed from (\ref{eq:RN_Sat_distance}) and (\ref{eq5}). Knowing $\vartheta(t)$, the actual Doppler shift can be straightforwardly compensated according to (\ref{eq:Doppler}). However, imagine that the assumed RN position is erroneous\footnote{Obviously, similar considerations hold true if the uncertainty concerns the position of the satellite.}. Bearing in mind Figure \ref{fig:true_dopp}, the assumed Doppler would be computed from the pair $(\left|\textbf{d}_e(t)\right|,\vartheta_e(t))$, which does not coincide with the real one, i.e., $(\left|\textbf{d}(t)\right|,\vartheta(t))$. With reference to Fig. \ref{fig:true_dopp}, we can relate $\vartheta_e(t)$ to $R_B$, which defines the ambiguity region, and $(\left|\textbf{d}(t)\right|,\vartheta(t))$ as 
\begin{equation}\begin{array}{l}
\left|\textbf{d}(t)\right|\text{cos}(\vartheta(t))=\left|\textbf{d}_e(t)\right|\text{cos}(\vartheta_e(t))+R_B\\
\left|\textbf{d}_e(t)\right|^2=\left|\textbf{d}(t)\right|^2+R^2_B-2R_B\left|\textbf{d}(t)\right|\text{cos}(\vartheta(t))
\label{eq8}\end{array}\end{equation}
Solving (\ref{eq8}), we end up with 
\begin{equation}
\text{cos}(\theta_e(t))=\frac{\left|\textbf{d}(t)\right|\text{cos}(\theta(t))-R_B}{\sqrt{\left|\textbf{d}(t)\right|^2+R^2_B-2R_B\left|\textbf{d}(t)\right|\text{cos}(\theta(t))}},
\end{equation}
which can be plugged into (\ref{eq:Doppler}). By performing an exhaustive search for different values of $R_B$ and $45^{\circ}$ $\leq \vartheta \leq 90^{\circ}$, as shown in Figure~\ref{fig:res_dopp}, it is possible to find the maximum position error that is tolerated such that the difference between the real and the assumed Doppler shift is below $950$ Hz. 

Therefore, for the considered scenario and the carrier frequency in the Ku-band, it is possible to conclude that the position error must be smaller than $4$ km. In this case, there is no need to modify the LTE waveform.

\subsection{Random Access and RN attach procedures}
In the considered Sat-RN architecture, the UEs in each on-ground cell perform the RA procedure with the corresponding RN, which terminates all protocols up to Layer 3. In particular, when implementing a contention-free RA, the only entities involved are the UE and the RN and, consequently, the delay on the satellite link is not involved and the procedure can be implemented without modifications. As for the contention-based RA, in Step 3 and 4 the RN shall contact the EPC, through the DeNB, so as to obtain a final network identifier for the UE. In this moment, the delay on the satellite channel is involved and shall be carefully taken into account. However, as previously reported, the contention timer in this phase of the RA procedure can be as large as $64$ ms, which is significantly above the RTT computed in Section~\ref{sec:Delay}. Thus, no modifications are required in the LTE RA procedure in order to implement it over a LEO satellite system.

With respect to the RN attach procedure, some difficulties might arise, similarly to the RA, due to the fact that the RTT is in the order of $20$ ms. In the RA procedure, as previously discussed, there are two separate timers to be taken into account: the RAR response window, as large as $15$ ms, and the contention resolution timer, which can be fixed to up to $64$ ms. The latter does not introduce any technical difficulty, as it is much larger than the RTT. However, the RAR response window, \emph{i.e.}, the period in which the RN is expecting a response to its RA request, is actually lower than the RTT. Anyway, it shall be noticed that this issue is present only at the RN start-up procedure. Since we are considering a system with fixed RNs (mobile RNs are not yet foreseen in the LTE standards) and the satellites in the LEO mega-constellation have known orbits, the RN attach procedure can be easily modified and implemented as an \emph{ad hoc} network deployment procedure, by exploiting all the available information.

\subsection{HARQ}
The critical aspect for HARQ in the considered system is: i) the time in which the RN can retransmit the data; and ii) the time window within which it is expecting a feedback (ACK/NACK) from the DeNB. Based on the above observations, these two parameters are $8$ ms and up to $7$ ms, respectively. As a consequence, it is not possible to use the standard HARQ procedure and proper solutions shall be found due to the large RTT over LEO satellites, due to the fact that the RTT computed in (\ref{eq:RTT_TwoWay}) is significantly larger than these values. In particular, the minimum number of parallel HARQ processes is defined as the ratio between the HARQ processing time, \emph{i.e.}, ACK time window plus propagation delay, and the Transmission Time Interval (TTI), \emph{i.e.}, $T_{HARQ}/TTI$. In the considered Sat-RN architecture, and assuming a $1$ ms TTI and an $8$ ms time window for the ACK reception, we have that:
\begin{equation}
\label{eq:HARQ_number}
    N_{HARQ} = \left\lceil \frac{T_{HARQ}}{TTI} = \frac{RTT + TTI}{TTI} \right\rceil = \left\lceil 21.06+1 \right\rceil = 23
\end{equation}
where $\left\lceil \cdot \right\rceil$ denotes the upper integer function. Such a large number of minimum HARQ parallel processes strongly impacts the UE buffer size, which shall be proportional to $N_{HARQ}\cdot TTI$, and the bit-width of the downlink control information fields, which now have to permit to discern among $23$ processes (minimum $5$ bits) instead of $8$ ($3$ bits) as in traditional LTE.

To cope with the above issue, several solutions can be envisaged: i) increasing the buffer size to cope with the large number of HARQ processes; ii) increase the number of HARQ processes, by maintaining the buffer size under control, by using a $2$ bit ACK to inform the transmitter on how close the received packet is to the originally transmitted one as currently under discussion for 5G systems \cite{3GPP_NR_ACK}. Therefore, the number of retransmission will be reduced, because the transmitter can add the redundant bits according to the feedback information; iii) reducing the number of HARQ processes and the buffer size, which also reduces the system throughput; and iv) not implementing the HARQ protocol, which requires solutions to solve issues related to colliding/non-decodable packets, such as transmitting repeated messages. All of these potential solutions need further investigation, in particular with respect to the impact on the overall system throughput.
\subsection{Satellite handover}
One last aspect to be carefully taken into account is related to the satellites' movement. In particular, satellites in a LEO mega-constellations orbit around the Earth with a large angular velocity. Thus, the RNs deployed on-ground have to cope with the fact that they will often be forced to to switch from the current satellite to another one in visibility as soon as the previous one falls behind the horizon or the minimum defined elevation angle. This is particularly critical since, in the LTE standard, the RNs are always assumed to be fixed and, thus, no mobility protocol or procedure is available for RNs that might need to change DeNB or link through which they are communicating with the EPC. To this aim, novel \emph{satellite handover} procedures need to be defined. In particular, two handover solutions are proposed hereafter:
\begin{enumerate}
    \item PHY-based handover: since the logical link between the RN and the DeNB is not aware of the physical channel being used for the communication, the best solution would be that of performing a PHY handover, \emph{i.e.}, an handover procedure limited to the PHY, which does not affect the upper layers. Such a handover procedure might pose significant challenges from the complexity point of view, in order to be seamless to the upper layers.
    \item Traditional handover: if the complexity of implementing a PHY-based handover is unacceptable, then the RN should perform a traditional handover. However, while complexity would be reduced, delays would be much larger. In particular, since for RNs the handover is not foreseen due to the absence of RN mobility in the LTE standard, the RN attach procedure shall start from scratch again at each handover, thus significantly reducing the system flexibility and throughput as, after start-up, all of the UEs in the cell shall perform a new RA.
\end{enumerate}
Both the above proposals pose technical challenges, either in terms of delays or of complexity, and thus need further analyses and assessment.

\section{Conclusions}
\label{sec:Conclusions}
In this paper, we focused on the definition of a LTE-based Satellite Communication system through LEO mega-constellations. In particular, two different architectures have been proposed to this aim: i) the Sat-RN architecture, in which the LTE on-ground cells are realised thanks to satellite-enabled Relay Nodes; and ii) the Sat-eNB architecture, in which the on-ground satellite-enabled entities are traditional LTE eNBs, with the additional capability to communicate through a satellite link. Focusing on former architecture, the impact of large Doppler shifts and delays has been discussed and assessed on the backhaul link in particular, while on the radio access link UEs do not perceive any difference with respect to a traditional terrestrial-only LTE cell. As for the Doppler shift, while the maximum value tolerated in LTE is $f_{d_{LTE}}=950$ Hz, the proposed analysis showed that, in the Ku-band $11<f_0<14$ GHz frequency range, the maximum Doppler shift experienced when the elevation of the satellite is $\vartheta=45^{\circ}$ is within this interval $158 \leq f_d \leq 201$ kHz. Therefore, the frequency offset is increased by a factor between $166$ and $221$ when compared to LTE. Consequently, it would be necessary to increase the robustness of LTE to be successfully used over a satellite in LEO as, for instance, by increasing the subcarrier spacing. An alternative solution might be that of allowing the receiver to further compensate the Doppler shift, by leveraging on the knowledge of its own coordinates and the orbit of the satellite. In this case, the study reveals that the position error must be smaller than $4$ km. With respect to the PHY/MAC procedures, the RN attachment can be replaced by ad-hoc network deployment, thanks to the knowledge of the (fixed) RNs locations and the satellites' orbit. As for the HARQ procedure, substantial modifications might be required, since the RTT is significantly larger than that used to design the LTE terrestrial procedure. In this case, several potential solutions have been proposed based on the configuration of the HARQ processes with different parameters, \emph{i.e.}, buffer sizes, number of parallel processes, and number of repetitions for the considered message. Finally, since relaying in LTE does not foresee mobility, a proper handover procedure shall be carefully designed in order to cope with the satellites' movement. The analysis and assessment of the proposed solutions, and, in particular, their impact on the overall system throughput and complexity, is left for further studies.

\section*{Acknowledgment}
This work has been supported by European Space Agency (ESA) funded activity SatNEx IV COO1-PART 2 WI 7 ``Air interface development for LEO constellation.'' The views of the authors of this paper do not necessarily reflect the views of ESA.

\nocite{*}% Show all bib entries - both cited and uncited; comment this line to view only cited bib entries;
%\bibliography{wileyNJD-AMA}%

\begin{thebibliography}{99}

\bibitem{ICC_paper}
    A. Guidotti, A. Vanelli-Coralli, M. Caus, J. Bas, G. Colavolpe, T. Foggi, S. Cioni, and A. Modenini, ``Satellite-enabled LTE systems in LEO Constellations,'' \emph{IEEE International Conference on Communications (ICC)}, Jun. 2017.

    \bibitem{3GPP36_201}
    3GPP TS 36.201 v14.4.0, ``Evolved Universal Terrestrial Radio Access (E-UTRA) and Evolved Universal Terrestrial Radio Access Network (E-UTRAN); Overall description; Stage 2 (Release 14),'' 2017.
    
    \bibitem{3GPP36_300}
    3GPP TS 36.300 v14.4.0, ``Evolved Universal Terrestrial Radio Access (E-UTRA) and Evolved Universal Terrestrial Radio Access Network (E-UTRAN); Overall description; Stage 2 (Release 14),'' 2017.
    
    \bibitem{HTS_1}
    H. Fenech, S. Amos, et. al, ``High throughput satellite systems: An analytical approach,'' \emph{IEEE Trans. on Aerospace}, vol. 51, no. 1, pp. 192--202, Jan. 2015.
    
    \bibitem{HTS_2}
    O. Vidal, et. al, ``Next generation High Throughput Satellite system,'' \emph{Proc. of IEEE ESTEL conference}, Oct. 2012.
    
    \bibitem{IntMgmTech1}
    P.-D. Arapoglou, A. Ginesi, S. Cioni, S. Erl, S. Andrenacci, A. Vanelli-Coralli, ``DVB-S2x Enabled Precoding for High Throughput Satellite Systems,'' \emph{Int. J. of Sat. Commun. and Net.}, vol. 34, pp. 439--455, Jun. 2015.
    
    \bibitem{IntMgmTech2}
    A. Ugolini, Y. Zanettini, A. Piemontese, A. Vanelli-Coralli, G. Colavolpe, ``Efficient satellite systems based on interference management and exploitation,'' \emph{Signals, Systems and Computers, 2016 50th Asilomar Conference on}, Nov. 2016.
    
    \bibitem{Evans2014}
    B. G. Evans, ``The Role of Satellites in 5G,'' \emph{7th Adv. Sat. Mult. Syst. Conf. and the 13th Sign. Proc. for Sp. Comm. Work. (ASMS/SPCS)}, 2014.

    \bibitem{Intro1}
    F. Bastia, C. Bersani, E. A. Candreva, S. Cioni, G. E. Corazza, M. Neri, C. Palestini, M. Papaleo, S. Rosati, and A. Vanelli-Coralli, ``LTE Adaptation for Mobile Broadband Satellite Networks,'' \emph{EURASIP J. on Wir. Comm. and Net.}, vol. 2009, Nov. 2009.

    \bibitem{Intro2}
    G. Araniti, M. Condoluci, and A. Petrolino, ``Efficient Resource Allocation for Multicast Transmissions in Satellite-LTE Networks,'' \emph{IEEE Glob. Comm. Conf. (GLOBECOM)}, Dec. 2013.

    \bibitem{Intro3}
    M. Amadeo, G. Araniti, A. Iera, and A. Molinaro, ``A Satellite-LTE network with delay-tolerant capabilities: design and performance evaluation,'' \emph{IEEE Vehic. Tech. Conf. (VTC Fall)}, Sep. 2011.
    
    \bibitem{Intro_Unibo}
    M. Papaleo, M. Neri, A. Vanelli-Coralli, and G. E. Corazza, ``Using LTE in 4G Satellite Communications: Increasing Time Diversity through Forced Retransmission,'' \emph{10th International Workshop on Signal Processing for Space Communications (SPSC)}, 2008.
    
    \bibitem{Intro4}
    H2020-ICT-2014-1 Project VITAL, Deliverable D2.3, ``System Architecture: Final Report,'' 2006.

    \bibitem{3GPP_5GSAT1}
    3GPP TSG RAN1, RP-171578, ``Propagation delay and Doppler in Non-Terrestrial Networks,'' 2017.
    
    \bibitem{3GPP_SAT_WI}
    3GPP TSG RAN1, RP-171450, Work Item description on ``Study on NR to support Non-Terrestrial Networks,'' Jun. 2017.

    \bibitem{3GPP_5GSAT2}
    3GPP TR 38.811 v0.2.0, ``Study on New Radio (NR) to support Non Terrestrial Networks (Release 15),'' 2017.
    
    \bibitem{Globecom_Unibo}
    O. Kodheli, A. Guidotti, and A. Vanelli-Coralli, ``Integration of Satellites in 5G through LEO Constellations,'' \emph{accepted to IEEE Globecom 2017}, Dec. 2017. available at: \url{https://arxiv.org/abs/1706.06013}
    
    \bibitem{3GPP_migration1}
    3GPP TS 36.801 v14.0.0, ``Study on new radio access technology: Radio access architecture and interfaces (Release 14),'' 2017.
    
    \bibitem{3GPP_migration2}
    3GPP TS 36.804 v14.0.0, ``Study on New Radio Access Technology; Radio Interface Protocol Aspects (Release 14),'' 2017.
        
    \bibitem{Propagation}
    C. Kourogiorgas and A. D. Panagopoulos, ``A Rain-Attenuation Stochastic Dynamic Model for LEO Satellite Systems Above 10 GHz,'' \emph{IEEE Trans. on Vehicular Tech.}, vol. 64, no. 2, pp. 829--834, Feb. 2015.

    \bibitem{3GPP36_401}
    3GPP TS 36.401 v14.0.0, ``Architecture description (Release 14),'' 2017.
    
    \bibitem{3GPP36_806}
    3GPP TS 36.806 v9.0.0, ``Relay architectures for E-UTRA (LTE-Advanced) (Release 9),'' 2010.
    
    \bibitem{3GPP36_116}
    3GPP TS 36.116 v14.0.0, ``Relay radio transmission and reception (Release 14),'' 2017.
    
     \bibitem{3GPP36_410}
    3GPP TS 36.410 v14.0.0, ``S1 general aspects and principles (Release 14),'' 2017.

    
    \bibitem{DVBS2X}
    ETSI EN 302 307-2 v1.1.1 (2015-02), ``Digital Video Broadcasting (DVB); Second generation framing structure, channel coding and modulation systems for Broadcasting, Interactive Services, News Gathering and other broadband satellite applications; Part 2: DVB-S2 Extensions (DVB-S2X),'' Feb. 2015.
    
    
    \bibitem{Doppler_LEO}
    I. Ali, N. Al-Dhahir, and J. E. Hershey, ``Doppler characterization for LEO satellites,'' \emph{IEEE Trans. on Commun.}, vol. 46, no. 3, pp. 309--313, 1998.

    \bibitem{Maral}
    G. Maral, M. Bousquets, and S. Zhili, ``Satellite Communications Systems: Systems, Techniques and Technology,'' Wiley 5th ed., 2009.

    \bibitem{3GPP36_211}
    3GPP TS 36.211 v14.4.0, ``Physical channels and modulation (Release 14),'' 2017.

    \bibitem{3GPP36_213}
    3GPP TS 36.213 v14.4.0, ``Physical layer procedures (Release 14),'' 2017.

    \bibitem{3GPP36_331}
    3GPP TS 36.331 v14.4.0, ``Radio Resource Control (RRC); Protocol specification (Release 14),'' 2017.

    \bibitem{3GPP_NR_ACK}
    3GPP TSG RAN1, R1-1704463, ``Considerations on CB grouping for multiple HARQ ACK/NACK bits per TB,'' 2017.


    
    
\end{thebibliography}

\end{document}